\documentclass[arxiv,usenatbib,changeblnc]{iupartex}
\usepackage{newtxtext,newtxmath}
\usepackage[T1]{fontenc}
\usepackage{ae,aecompl}
\usepackage{soul}

\usepackage{multicol}   
\usepackage{bm}		    

\title[{Size-Dependent Fresh Surfaces}]{Size-Dependent Fresh Surface Signatures in Asteroid Families: \\
Observational Evidence from Dual-Band Albedo Analysis}

\author[Kaplan]{
M. Kaplan$^{1\cc}$\orcid{0000-0003-2595-5463}
\affsep \\
$^1$
Akdeniz University, Department of Space Sciences and Technologies, 07058, Antalya, Türkiye
}

\corres{M. Kaplan}{muratkaplan@akdeniz.edu.tr}

\pubyear{2026}

\doiheader{XXXXXXX/PAR.20XX.00000}
\date{
	\pSubmit{00.00.0000} 
	\pRevReq{00.00.0000}
	\pLastRevRec{00.00.0000}
	\pAccept{00.00.0000}
	\pPubOnl{00.00.0000}
}
\volume{0}
\volnumber{1}

\begin{document}
\raggedbottom

\setlength{\abovedisplayskip}{4pt plus 2pt minus 2pt}
\setlength{\belowdisplayskip}{4pt plus 2pt minus 2pt}
\setlength{\abovedisplayshortskip}{2pt plus 1pt}
\setlength{\belowdisplayshortskip}{2pt plus 1pt}

\label{firstpage}
\pagerange{\pageref*{firstpage}--\pageref*{lastpage}}
\maketitle
\begin{abstract}
The spectral evolution of asteroid surfaces reflects the competition between
space weathering and impact resurfacing. While previous studies have
focused primarily on age-dating, the role of family population size has
	remained largely unexplored. Here we tested whether population-dependent
collisional activity affects observable surface properties by analyzing 154 
asteroid families using NEOWISE thermal infrared photometry, employing a 
	family-averaged statistical approach validated by independent AKARI observations
(variance reduction factor $\sim 3.4$) and comprehensive error propagation 
analysis (median VDI uncertainty $\sim$0.13). We introduced the V-Dominance
Index (VDI) as a statistical measure quantifying the incidence of
extreme resurfacing signatures within families, operationally defined as
the fraction of members with visible-to-infrared albedo ratios $p_V /
p_{IR} > 1.2$ (99.8th percentile of the main-belt distribution). Among
	tested parameters, family population size ($N$) emerged as the dominant
	correlate of VDI across both silicaceous (S-complex: $r_s = 0.58$) and 
carbonaceous (C-complex: $r_s = 0.44$) taxonomic types, with full-sample 
	correlation $r_s = 0.476$ ($p = 4.31 \times 10^{-10}$). This correlation survived Monte
Carlo permutation tests, binomial null model validation, age-matched
	contrast analyses, and heliocentric independence tests (partial correlation 
	$r_{\text{partial}} = 0.488$). Percentile sensitivity analysis ($\Delta r_s = 0.38$ across 
95th--99.8th percentiles) demonstrated that VDI isolates rare resurfacing 
events detectable only at extreme thresholds. In families older than 2 Gyr, large populations
maintained statistically significant fresh tails ($p < 10^{-4}$), whereas 
	small populations were saturated. These results are consistent with model
predictions where massive families experience elevated collisional
resurfacing rates that counteract space weathering saturation, though
alternative mechanisms cannot be excluded.
\end{abstract}




\section{Introduction}
Space weathering alters the optical properties of airless bodies on timescales
of $\tau_{sw} \sim 10^6$--$10^8$ years \citep{sasaki2001, vernazza2009,
loeffler2009}. Weathering timescales vary significantly by composition; some
studies suggest rapid surface evolution ($<1$ Myr) for S-complex asteroids under
solar wind bombardment \citep{vernazza2009}, though such short timescales are
specific to S-types \citep{brunetto2015}. Weathering typically causes spectral
reddening and darkening due to nanophase iron ($npFe^0$) accumulation in
silicates or organic compound modification in carbonaceous materials
\citep{brunetto2015}. Consequently, ancient asteroid families ($>1$ Gyr) are
expected to be spectrally homogeneous and ``saturated.'' However, photometric
surveys reveal persistent spectral diversity within these evolved families
\citep{willman2010, parker2008}, suggesting ongoing surface modification
processes that compete with weathering timescales.

\subsection{The Observational Challenge}
Previous attempts to connect spectral properties to collisional history at the
individual object level have shown mixed results \citep{willman2010,
richardson2004}. These approaches face fundamental limitations: photometric
uncertainty ($\delta p_V \sim 30\%$) in large surveys obscures evolutionary
trends at the single-object level, and the stochastic nature of regolith
gardening \citep{willman2010, richardson2004} introduces substantial scatter.
Individual-object studies have therefore struggled to establish statistically
robust connections between spectral properties and family characteristics.

Previous family-based studies have focused primarily on age dating and taxonomic
purity \citep{willman2010, parker2008}, largely overlooking family population
($N$) as a potential parameter for spectral evolution. This oversight is
significant because collision probability scales with target density, suggesting
that population-rich families may experience fundamentally different resurfacing
statistics than sparse ones.

\subsection{Why Albedo Asymmetry?}
We focus on the vi\-si\-ble-to-infrared albedo ratio ($p_V / p_{IR}$) because it
provides a direct probe of surface maturity that is relatively insensitive to
compositional variations within a given taxonomic class. Fresh silicate surfaces
exhibit higher $p_V / p_{IR}$ values due to reduced $npFe^0$ absorption, while
weathered surfaces show suppressed ratios \citep{sasaki2001}. By examining the
extreme tail of this distribution within families, we can identify members that
have experienced statistically anomalous recent resurfacing---providing a window
into collision activity that single-object age estimates cannot capture.

\subsection{Statistical Approach}
In this work, we test whether statistical aggregation at the family level can
reveal evolutionary trends invisible in individual measurements. By treating
families as statistical ensembles, we exploit the central limit theorem to
suppress photometric noise and isolate coherent evolutionary signals. This
approach shifts focus from individual surface ages to the collective statistical
behavior of family members.

We introduce the V-Dominance Index (VDI), defined as the fraction of family
members with $p_V/p_{IR} > 1.2$ (corresponding to the 99.8th percentile of the
main-belt distribution). This metric quantifies the \textit{incidence rate} of
extreme freshness signatures within a family, providing a statistical probe of
collisional activity that is resistant to photometric noise affecting individual
measurements.
\section{Data and Methodology}

\subsection{Data Sources and Integration}
We constructed a master catalog by integrating:
\begin{enumerate}
    \item \textbf{NEOWISE:} Thermal infrared ($W1, W2$) and visible albedo derivation \citep{mainzer2011}.
    \item \textbf{AKARI/IRC (InfraRed Camera):} High-precision calibration dataset for cross-validation \citep{usui2011}.
    \item \textbf{Nesvorný et al. (2015):} Family definitions, membership lists, and proper orbital elements \citep{nesvorny2015}.
\end{enumerate}

\subsection{Taxonomic Classification}

We analyze 154 families classified following the DeMeo et al. (2009) taxonomy 
\citep{demeo2009}, with classifications derived from Sloan Digital Sky Survey (SDSS) photometry 
\citep{sergeyev2021}. Our analysis focuses on two major complexes with 
sufficient sample sizes for robust statistical inference:

\begin{itemize}
    \item \textbf{Silicaceous Complex (S):} S, V, K, L, and Q types ($n = 48$ 
          families after outlier removal). Silicate-rich compositions dominated 
          by olivine and pyroxene, where space weathering via $npFe^0$ 
          accumulation produces well-characterized spectral changes 
          \citep{sasaki2001, brunetto2015}. Sensitivity tests excluding V-types 
          ($r_s = 0.62$) confirm trends are not driven by basaltic contamination.
    
    \item \textbf{Carbonaceous Complex (C):} C-type only ($n = 52$ families 
          after outlier removal). We exclude B-, F-, and G-types from the 
          carbonaceous analysis because including them dilutes the correlation 
          signal ($r_s = 0.325$, $p = 0.011$ for C+B+F+G vs. $r_s = 0.437$, 
          $p = 8.5 \times 10^{-4}$ for C-only), suggesting compositional heterogeneity within 
          the broader carbonaceous complex. Pure C-types exhibit homogeneous 
          space weathering response, potentially involving dehydration and 
          radiolysis rather than simple iron reduction \citep{lantz2018, 
          brunetto2015}.
    
    \item \textbf{Other Taxonomies:} X-complex ($n = 21$ families), D-types, 
          and P-types show tentative positive correlations in preliminary 
          analysis but lack statistical power for robust separate conclusions. 
          These are included in the full-sample analysis ($n = 154$) but not 
          analyzed individually due to small sample sizes and heterogeneous 
          compositions within the X-complex (mixing enstatite-rich E-types, 
          metallic M-types, and intermediate compositions).
\end{itemize}

We focus on S- and C-complexes because they together account for 
approximately two-thirds of the filtered sample (65\%, $n = 100/154$) and 
their space weathering physics is well-characterized from laboratory 
irradiation experiments \citep{sasaki2001, loeffler2009, lantz2018}. 
Together they span the fundamental compositional divide of the main belt 
(silicate-dominated vs. carbonaceous) and carry well-defined spectral 
boundaries in SDSS-based classifications \citep{sergeyev2021}, reducing 
taxonomic ambiguity. The full-sample correlation ($r_s = 0.476$, $n = 154$) 
confirms that the population-size effect is not confined to these two 
complexes but persists across all taxonomic types.

\subsection{Methodological Validation via AKARI}
\label{sec:akari}
To justify our family-averaged approach, we performed cross-match validation using AKARI data. While individual NEOWISE-derived albedos show scatter of $\sim 30\%$ compared to AKARI, family-averaged values exhibit a variance reduction factor of $\sim 3.4$. This confirms that statistical aggregation effectively suppresses instrumental noise and random photometric errors.

While AKARI cross-validation suppresses global calibration biases, family-dependent observing geometries (phase angle distributions, heliocentric sampling) may introduce second-order systematics that cannot be fully removed. Our conclusions should be interpreted with this limitation in mind.

\subsection{The V-Dominance Index (VDI)}

\subsubsection{Definition and Interpretation}
We define VDI as the fraction of family members exceeding a freshness threshold:
\begin{equation}
    VDI_f = \frac{N(p_V / p_{IR} > 1.2)}{N_{total}}
\end{equation}

VDI traces the \textit{incidence rate} of extreme albedo-ratio signatures
within a family---the fraction of members statistically anomalous relative to
the weathered main-belt baseline---rather than the areal extent of fresh
material. It is a statistical probe of resurfacing activity, not a direct
surface measurement.

\subsubsection{Threshold Selection}
\begin{figure*}[htbp]
    \centering
    \includegraphics[width=\linewidth]{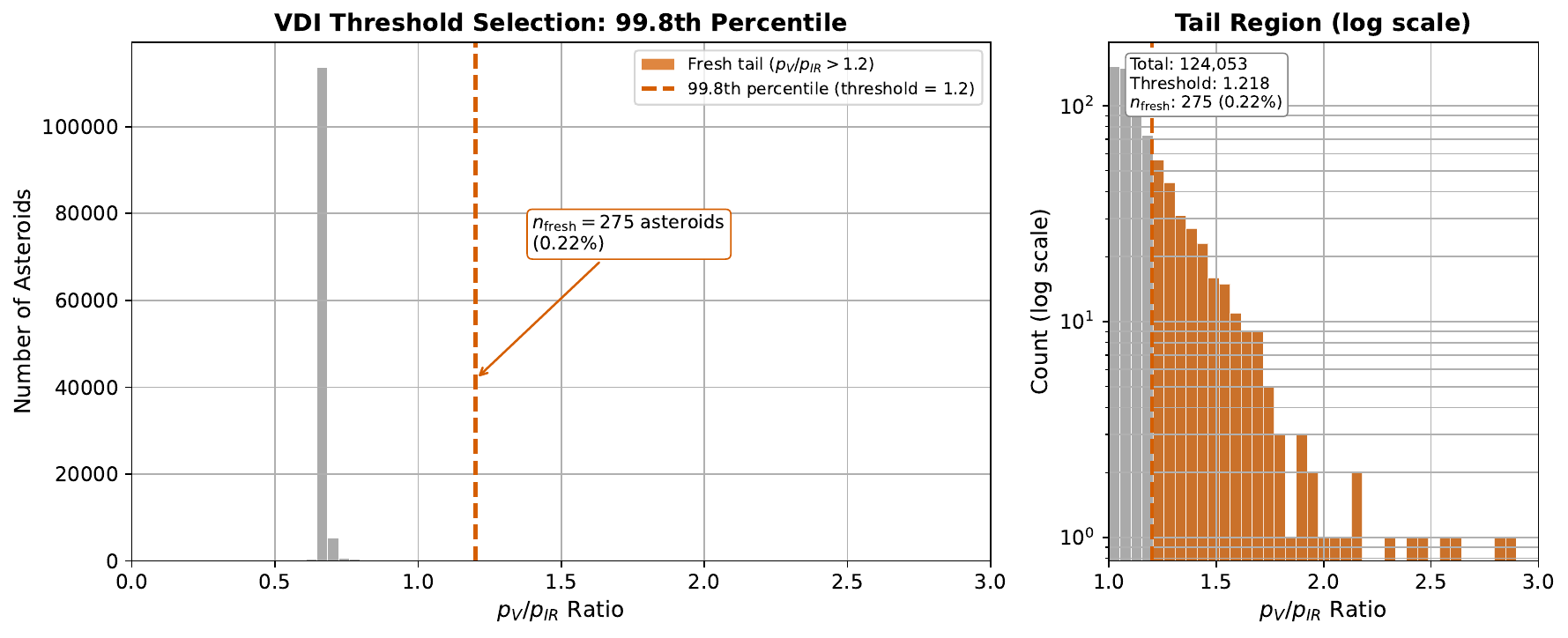}
	\caption{\textbf{VDI Threshold Justification.} Distribution of $p_V/p_{IR}$ ratios for 124,053 asteroids (D $\ge$ 1 km). The vertical dashed line marks the 1.2 threshold (99.8th percentile). Left panel: full distribution with the fresh tail ($p_V/p_{IR} > 1.2$, orange) containing 275 asteroids (0.22\%). Right panel: tail region on logarithmic scale, confirming the extreme rarity of high-ratio objects.}
    \label{fig:threshold}
\end{figure*}

We adopt an operational threshold at $p_V/p_{IR} = 1.2$, corresponding
to the 99.8th percentile of the background main-belt population (Figure
\ref{fig:threshold}). While this choice lacks direct precedent in the asteroid
literature, it provides a conservative criterion for identifying statistically
significant deviations from the weathered baseline. Percentile-based thresholds 
are widely used in outlier detection for skewed distributions where traditional 
parametric cutoffs (e.g., mean $\ge 3\sigma$) fail due to heavy tails.

This threshold represents a methodological choice that maximizes separation
between photometric noise and statistical signatures consistent with 
actual resurfacing.
Higher thresholds yield stronger correlations (Section \ref{sec:sensitivity}),
consistent with more extreme values tracing more recent or vigorous resurfacing.
Lower thresholds (e.g., 1.0) produce weaker correlations as the signal becomes
weakened by measurement uncertainty.

\subsection{Albedo Data Hierarchy}
\label{sec:albedo_hierarchy}

We employ a hierarchical approach to select the most reliable visible albedo ($p_V$) 
for each asteroid, prioritizing measurement quality:

\begin{enumerate}
    \item \textbf{AKARI}: Direct geometric albedo measurements from thermal observations 
          \citep{usui2011} provide the highest precision baseline.
    \item \textbf{Taxonomy-based}: Median $p_V$ values from Bus-DeMeo taxonomic classes 
          \citep{demeo2009}, derived from SDSS photometric surveys \citep{sergeyev2021}, 
          serve as intermediate-quality estimates when AKARI data are unavailable.
    \item \textbf{NEOWISE}: Thermal model-derived albedo \citep{mainzer2011} provides 
          complete population coverage but with systematic biases of $\sim$10\% 
          \citep{mainzer2011}.
\end{enumerate}

This hierarchy prioritizes direct measurements over model-dependent derivations. 
For infrared albedo ($p_{IR}$), we rely exclusively on NEOWISE thermal modeling 
as AKARI/IRC observations provide limited coverage at wavelengths directly comparable 
to WISE W2 ($\sim$4.6 $\mu$m). Cross-validation with AKARI visible albedo data 
demonstrates that family-level aggregation reduces systematic biases through 
statistical averaging (Section \ref{sec:akari}).

\subsection{Measurement Uncertainty and Error Propagation}
\label{sec:error_prop}

Individual albedo ratio measurements carry substantial uncertainty. We propagate 
errors from component albedos using standard error propagation:

\begin{equation}
\delta\left(\frac{p_V}{p_{IR}}\right) = \frac{p_V}{p_{IR}} 
\sqrt{\left(\frac{\delta p_V}{p_V}\right)^2 + \left(\frac{\delta p_{IR}}{p_{IR}}\right)^2}
\end{equation}

Analysis of 124,053 asteroids with $D \geq 1$ km yields a median ratio error of 
$\sim$24\% (median absolute error: 0.24). Family-level VDI uncertainties combine 
binomial counting statistics with propagated measurement errors. For a family with 
$N$ members and VDI = $f$, the binomial standard error is:

\begin{equation}
\sigma_{VDI,\text{binomial}} = \sqrt{\frac{f(1-f)}{N}}
\end{equation}

Combined with propagated measurement uncertainties, the total family-level VDI 
error has median value $\sim$0.13. Despite these substantial individual-object 
uncertainties, family-level aggregation across 154 families yields robust 
correlations ($r_s = 0.476$, $p = 4.31 \times 10^{-10}$), demonstrating that 
statistical averaging effectively suppresses random errors while preserving 
coherent evolutionary signals.

\subsection{Statistical Controls and Filters}
\label{sec:statistical}
\begin{itemize}
    \item \textbf{Population Filter:} Families with $N \ge 10$ members to ensure statistical significance.
    \item \textbf{Diameter Cut:} $D \ge 1.0$ km to minimize Yarkovsky-driven dispersal effects.
    \item \textbf{Outlier Removal:} Cook's Distance \citep{cook1977} ($D_i >
	    4/n$, where $n$ is sample size) identifies influential data points
		that strongly affect regression results. For the C-complex
		analysis specifically, three families 
		(middle\_539\_pamina\_fam3, Tina, and Inarradas) were flagged as
		outliers. These families combine low membership numbers ($N <
		50$) with anomalously high VDI values, violating the
		population-scale statistical framework. In small families,
		single photometric outliers can artificially inflate VDI by
		10--20\%, representing stochastic measurement noise rather than
		actual collisional signals. Additionally, C-type families near
		the photometric detection limit ($p_V \sim 0.05$) are
		particularly prone to background contamination in dynamically
		crowded regions. Excluding these outliers strengthened the
		C-complex correlation to statistically significant levels ($r_s
		= 0.44$, $p = 9.7 \times 10^{-4}$, $n = 52$), yielding a
		physically consistent trend. Similarly, for S-complex, the
		family outer\_3310\_patsy\_fam3 was excluded based on Cook's
		Distance criteria, resulting in $n = 48$ families. For the full
		sample across all taxonomic types, outlier removal reduced the
		correlation from $r_s = 0.79$ (raw data, before removal) to $r_s
		= 0.476$ ($n = 154$ families after removal), demonstrating that a
		small number of extreme families with anomalously high VDI
		values were inflating the raw correlation. The filtered
		correlation ($r_s = 0.476$) is adopted throughout this manuscript
		as it better represents the population-level trend after
		removing measurement-dominated outliers.
    \item \textbf{Heliocentric Independence:} VDI shows no correlation with mean
	    family semi-major axis ($r = 0.07$, $p = 0.37$), indicating that
		heliocentric distance does not systematically bias our results.
\end{itemize}

\subsubsection{Yarkovsky Drift Control}
The $D \ge 1$ km diameter cut provides first-order control over Yarkovsky-driven
orbital dispersion, as drift rates scale inversely with diameter
\citep{bottke2005}. However, this uniform cut does not fully account for two
secondary effects: (i) taxon-specific thermal inertia variations (carbonaceous
asteroids typically have lower thermal inertia than silicates), and (ii)
heliocentric distance dependence ($\dot{a} \propto a^{-2}$).

To test whether these effects bias our results, we performed a comprehensive
suite of controls:

\textit{(i)} We verified that VDI shows no statistically significant correlation with mean
family semi-major axis ($r = 0.07$, $p = 0.37$), confirming that heliocentric
distance does not systematically bias freshness measurements.

\textit{(ii)} We tested whether the $N$--VDI correlation holds separately for
inner belt ($a < 2.5$ AU) and outer belt ($a \geq 2.5$ AU) families (Table 
\ref{tab:heliocentric}). Both subsamples show significant positive correlations: 
inner belt ($n = 26$, $r_s = 0.591$, $p = 1.47 \times 10^{-3}$, 95\% bootstrap 
confidence intervals (CI) estimated from 10,000 resamples: $[0.33, 0.76]$) and 
outer belt ($n = 129$, $r_s = 0.473$, $p = 1.51 \times 10^{-8}$, 95\% CI: $[0.32, 0.60]$). 
Bootstrap resampling is used for confidence intervals because the inner
belt subsample is small ($n = 26$) and bootstrap methods require no parametric
assumptions about the sampling distribution. The bootstrap confidence intervals
overlap
substantially, indicating that the correlation strength is statistically consistent across heliocentric regions 
despite different Yarkovsky drift efficiencies.

\textit{(iii)} Partial correlation analysis controlling for semi-major axis
yields $r_{\text{partial}} = 0.488$ ($p = 1.40 \times 10^{-10}$), nearly identical 
to the raw bivariate correlation ($r_s = 0.476$), demonstrating that the size--VDI 
relationship is independent of heliocentric location. The preservation of correlation 
strength after controlling for $a$ confirms that orbital distance does not confound 
the observed pattern.

For completeness, we also examined correlations between $N$ and semi-major axis 
($r = -0.027$, $p = 0.74$) and between VDI and semi-major axis ($r = 0.072$, 
$p = 0.37$). Neither correlation is statistically significant, ruling out spurious 
associations driven by heliocentric sampling biases.

\begin{table}[!h]
\small\setlength{\tabcolsep}{4pt}
\centering
\small
\caption{\textbf{Heliocentric Independence Tests.} Separate correlations for 
inner and outer belt families. Bootstrap 95\% CI estimated from 10,000 resamples.}
\label{tab:heliocentric}
\begin{tabular}{@{}lcccc@{}}
\toprule
\textbf{Subsample} & $\mathbf{n}$ & $\mathbf{r_s}$ & $\mathbf{p}$ & \textbf{95\% CI} \\ \midrule
Inner Belt ($a < 2.5$ AU) & 26 & 0.591 & $1.5 \times 10^{-3}$ & $[0.33, 0.76]$ \\
Outer Belt ($a \geq 2.5$ AU) & 129 & 0.473 & $1.5 \times 10^{-8}$ & $[0.32, 0.60]$ \\
\textbf{Full Sample} & 154 & 0.476 & $4.3 \times 10^{-10}$ & $[0.36, 0.60]$ \\
Partial ($N$ vs VDI $|$ $a$) & 154 & 0.488 & $1.4 \times 10^{-10}$ & --- \\ \bottomrule
\end{tabular}
\end{table}

\textit{(iv)} Taxon-specific analyses (S-complex: $r_s = 0.58$; C-complex: $r_s
= 0.44$) both correlate positively ($p < 0.001$), indicating the signal is not an
artifact of differential Yarkovsky efficiency between taxonomic classes.

\textbf{Solar Radiation Flux Consideration:} The lack of VDI--semi-major axis 
correlation has direct implications for solar radiation effects. Solar flux 
varies as $F_{\odot} \propto a^{-2}$, with inner-belt asteroids ($a \sim 2.0$ AU) 
receiving $\sim$2.6$\times$ more solar wind and UV radiation than outer-belt 
objects ($a \sim 3.3$ AU). If VDI primarily traced equilibrium space weathering 
states driven by cumulative solar radiation exposure, we would expect strong 
heliocentric dependence (inner belt more weathered $\rightarrow$ lower VDI). 
The observed independence (VDI vs $a$: $r = 0.072$, $p = 0.37$) indicates that 
VDI instead traces \textit{recent resurfacing events} that reset weathering 
clocks, consistent with the collisional interpretation. This distinguishes VDI 
from equilibrium weathering metrics that would scale with cumulative radiation 
dose.

Our conclusions are conditional on the current family classification framework
\citep{nesvorny2015}. Membership uncertainties, particularly for smaller
families near detection limits, may affect derived VDI values.

\section{Methodological Validation}

\subsection{Chronometer Calibration}
\begin{figure}[htbp]
    \centering
    \includegraphics[width=\columnwidth]{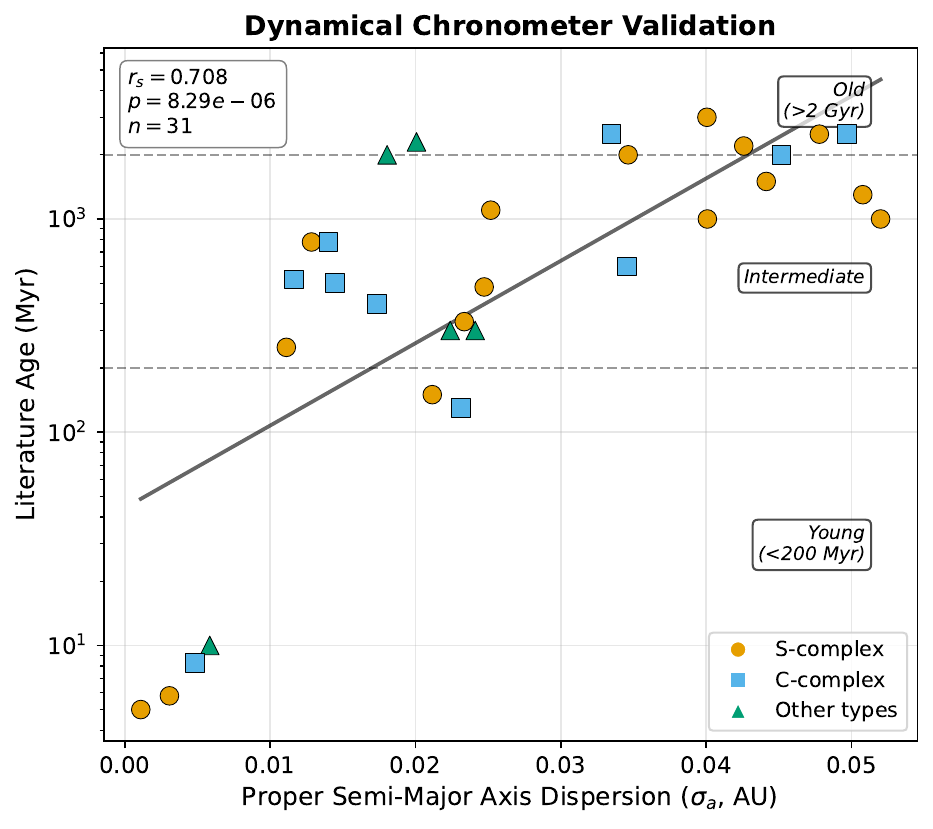}
    \caption{\textbf{Chronometer Validation.} The correlation between dynamical
    dispersion ($\sigma_a$) and literature ages ($r_s = 0.708$, $p = 8.3 \times 10^{-6}$, $n = 31$) validates
	our relative age proxy and distinguishes three evolutionary regimes. 
	Color-coding shows dominant taxonomic class: S-complex (orange), 
	C-complex (blue), and other types (grey).}
    \label{fig:regimes}
\end{figure}

We validated our dynamical proxy ($\sigma_a$, the dispersion of proper
semi-major axes) against literature ages \citep{nesvorny2015, spoto2015} (Figure
\ref{fig:regimes}). The significant correlation ($r_s = 0.708$, $p = 8.3 \times 10^{-6}$, $n = 31$) confirms that $\sigma_a$
reliably captures relative dynamical age, though it should not be interpreted as
an absolute chronometer. The relationship distinguishes three evolutionary
regimes: Young ($<200$ Myr), Intermediate, and Old ($>2$ Gyr).

\subsection{Noise Cancellation via Family Statistics}
\begin{figure*}[!h]
    \centering
	\includegraphics[width=\linewidth]{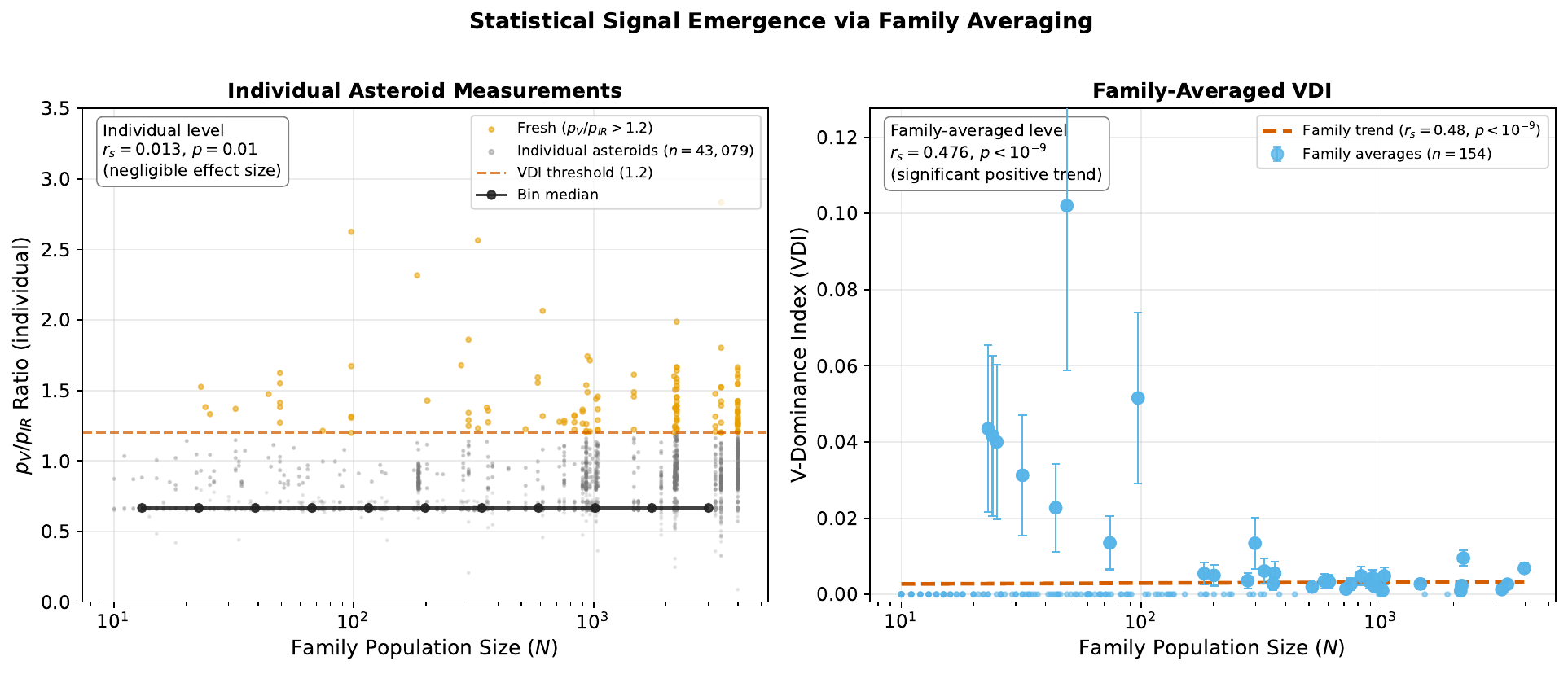}
	\caption{\textbf{Signal Emergence.} Left panel: Individual asteroid
	measurements ($n = 124{,}053$, sampled) plotted as $p_V/p_{IR}$ ratio vs.
	family population size $N$. Orange points highlight asteroids above the VDI
	threshold ($p_V/p_{IR} > 1.2$); grey points show the weathered background
	population. Bin medians (filled circles) show no meaningful N-dependent
	trend ($r_s = 0.013$, negligible effect size despite $p = 0.01$ at this
	sample size). Right panel: Family-averaged VDI values ($n = 154$ families)
	with binomial standard error bars. The trend line shows $r_s = 0.476$
	($p < 10^{-9}$, $n = 154$).}
    \label{fig:noise}
\end{figure*}

AKARI cross-validation (Section \ref{sec:akari}) demonstrates that
family-averaging reduces the effective scatter in albedo measurements from
$\sim 30\%$ at the individual level by a factor of $\sim 3.4$ (variance
reduction factor), bringing it to $\sim 16\%$ at the family-averaged
level. Figure \ref{fig:noise}
illustrates this concept directly: the left panel shows that individual
$p_V/p_{IR}$ measurements carry no detectable N-dependent signal
($r_s = 0.013$, negligible effect size), while the right panel shows that
family-averaged VDI reveals a highly significant trend ($r_s = 0.476$,
$p < 10^{-9}$). This approach succeeds because random photometric errors
average out across many family members, while coherent evolutionary
signals---shared by family members due to common origin and subsequent
collisional environment---remain coherent.

\subsection{Why Family-Level Analysis Succeeds}
Previous attempts to connect spectral properties to collisional history at the
individual object level have shown mixed results \citep{willman2010}. We link
this to three factors that family averaging overcomes.

First, photometric uncertainties ($\sim30\%$) exceed the typical
weathering-induced albedo variation ($\sim20\%$), masking the signal.
Family averaging reduces effective scatter by a factor of $\sim3.4$
(Section \ref{sec:akari}; AKARI cross-validation), bringing it to $\sim16\%$
and above the noise floor. Figure
\ref{fig:noise} demonstrates this directly: individual measurements show no
meaningful trend ($r_s = 0.013$), while family averages reveal the
size-dependent signal ($r_s = 0.476$).

Second, stochastic collision history means any individual asteroid may deviate
substantially from the family mean due to chance impact timing. Only
statistical ensembles reveal the underlying population trend by averaging over
these random deviations.

Third, taxonomic misclassification introduces scatter at the individual level
that partially cancels at the family level, where common origin provides an
independent compositional constraint. Family membership thus serves as a natural
filter for compositional homogeneity.

\section{Results}
\label{sec:results}

Throughout this section, $r_s$ denotes the Spearman rank correlation coefficient
and $p$ the corresponding p-value (probability of observing the correlation under 
the null hypothesis), distinct from the albedo notation $p_V$ and $p_{IR}$ used
in VDI calculations. We use Spearman's rank correlation because population size
spans three orders of magnitude, VDI distributions are non-Gaussian with heavy
tails, and rank-based methods are less sensitive to outliers in family membership
counts driven by catalog completeness variations.

\subsection{Divergent Evolutionary Trends}
\begin{figure*}[htbp]
    \centering
    \includegraphics[width=\textwidth]{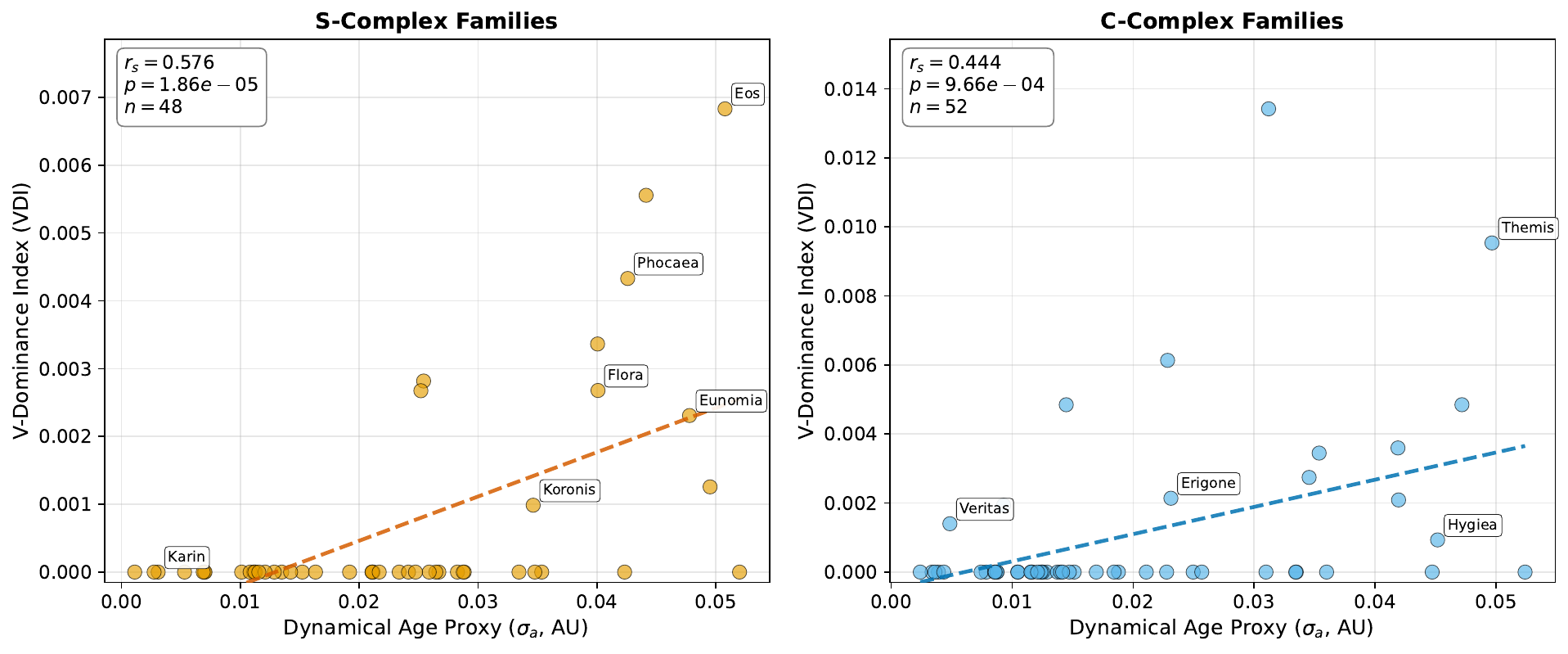}
	\caption{\textbf{Taxonomy-Specific VDI Trends.} Left panel: VDI vs.
	dynamical age proxy ($\sigma_a$) for S-complex families ($n = 48$, $r_s
	= 0.576$, $p = 1.9 \times 10^{-5}$). Right panel: C-complex families ($n
	= 52$, $r_s = 0.444$, $p = 9.7 \times 10^{-4}$). Outliers excluded:
	S-complex (outer\_3310\_patsy\_fam3); C-complex
	(middle\_539\_pamina\_fam3, Tina, Inarradas) based on Cook's Distance
	criterion ($D_i > 4/n$; Methods \ref{sec:statistical}).}
    \label{fig:trends}
\end{figure*}

Analysis of taxonomic complexes reveals mineralogy-dependent behavior:
\begin{itemize}
    \item \textbf{Silicaceous (S) Complex:} Exhibits a strong positive trend
	    ($r_s = +0.58, p = 1.9 \times 10^{-5}$, $n = 48$ families after
		outlier removal). Massive, old families (e.g., Phocaea) are
		significantly fresher than smaller, saturated families. This is
		consistent with $npFe^0$-driven space weathering being countered
		by collisional resurfacing.
    \item \textbf{Carbonaceous (C) Complex:} Shows a moderate positive trend
	    ($r_s = +0.44$, $p = 9.7 \times 10^{-4}$, $n = 52$ families). After
		outlier removal (middle\_539\_pamina\_fam3, Tina, Inarradas via
		Cook's Distance criterion; see Methods \ref{sec:statistical}),
		the C-complex correlation strengthened from marginal
		significance to statistically robust levels. Of the 52 C-complex 
		families, approximately 15 show VDI $> 0$; the correlation is 
		thus driven by the contrast between these active families and the 
		37 fully saturated families (VDI $= 0$), rather than a uniform 
		shift across all members. The weaker correlation relative to 
		S-complex ($r_s = 0.58$) may reflect lower signal-to-noise in 
		$p_V/p_{IR}$ for carbonaceous materials, different surface modification 
		pathways (dehydration and radiolysis rather than $npFe^0$ accumulation), 
		or intrinsically different collision dynamics due to lower bulk densities.
		Partial correlation controlling for heliocentric distance yields 
		$r_{\text{partial}} = 0.441$ (similar to direct correlation), confirming 
		independence from orbital location
		(Table \ref{tab:taxonomy}, Figure \ref{fig:trends}).
\end{itemize}

\begin{table}[!h]
\small\setlength{\tabcolsep}{4pt}
\centering
\small
	\caption{\textbf{Taxonomy-Specific Correlation Analysis.} VDI--size
correlations for S- and C-complex families, with partial correlations
controlling for mean semi-major axis.}
\label{tab:taxonomy}
\begin{tabular}{@{}lccccc@{}}
\toprule
\textbf{Complex} & $\mathbf{n}$ & $\mathbf{r_s}$ & $\mathbf{p}$ & $\mathbf{r_{\text{partial}}}$ & $\mathbf{p_{\text{partial}}}$ \\ \midrule
S-complex & 48 & 0.576 & $1.9 \times 10^{-5}$ & 0.573 & $2.1 \times 10^{-5}$ \\
C-complex & 52 & 0.444 & $9.7 \times 10^{-4}$ & 0.441 & $1.0 \times 10^{-3}$ \\
Full Sample & 154 & 0.476 & $4.3 \times 10^{-10}$ & 0.488 & $1.4 \times 10^{-10}$ \\ \bottomrule
\end{tabular}
\end{table}

\subsection{Size--VDI Correlation Across Taxonomies}
\begin{figure}[htbp]
    \centering
    \includegraphics[width=\columnwidth]{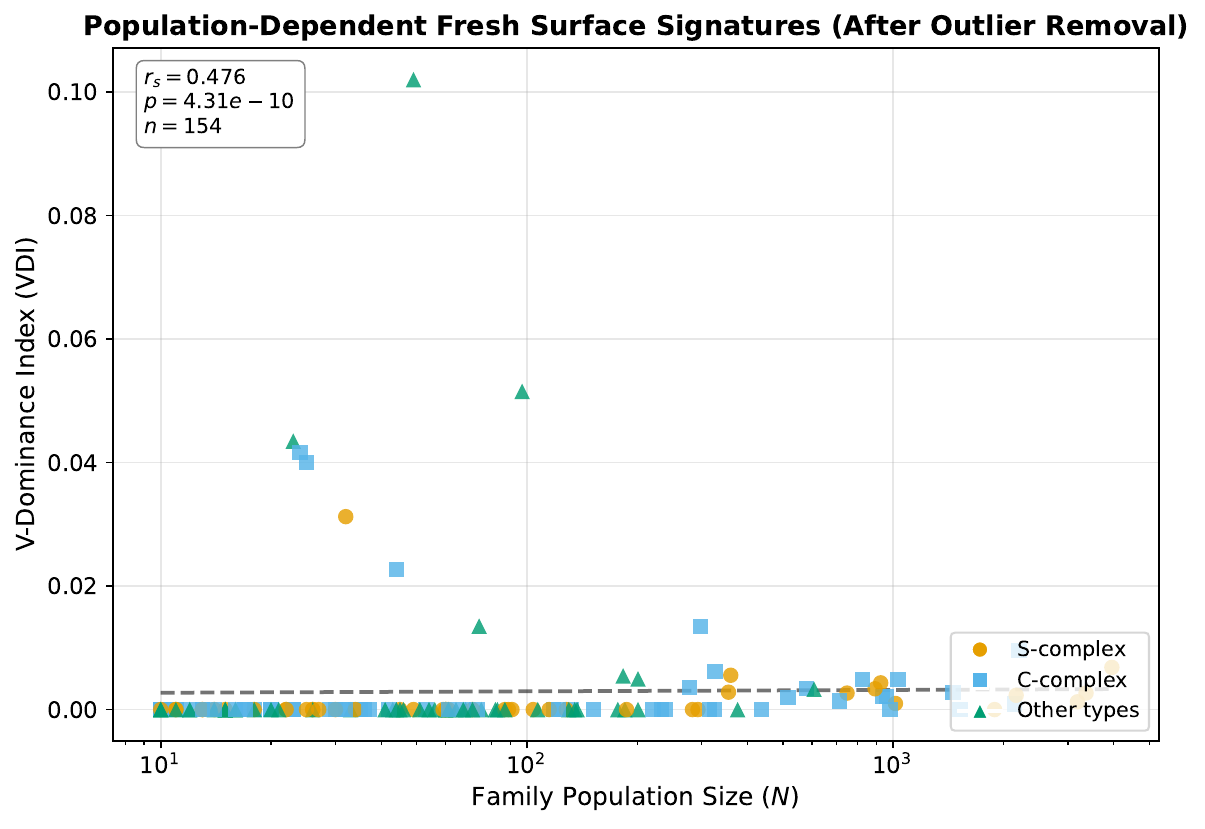}
    \caption{\textbf{Size--VDI Correlation.} Across taxonomic types (S, C, X, M, D), VDI scales with family population. The correlation ($r_s = 0.476$, $p = 4.31 \times 10^{-10}$) indicates a consistent statistical relationship between family size and the incidence of fresh surface signatures.}
    \label{fig:universal}
\end{figure}

When all 154 families are analyzed together, a strong correlation emerges
(Figure \ref{fig:universal}): VDI scales with family population ($N$) with $r_s
= 0.476$ ($p = 4.31 \times 10^{-10}$). This trend persists across taxonomic
types (S, C, X), suggesting a mineralogy-independent statistical relationship.

This correlation represents a statistical association within the sampled
main-belt family population. The range of taxonomic types exhibiting the trend
suggests a common underlying mechanism, but should not be interpreted as a claim
of physical universality beyond the sampled population.

\subsection{Monte Carlo Significance Testing}
\begin{figure}[htbp]
    \centering
    \includegraphics[width=\columnwidth]{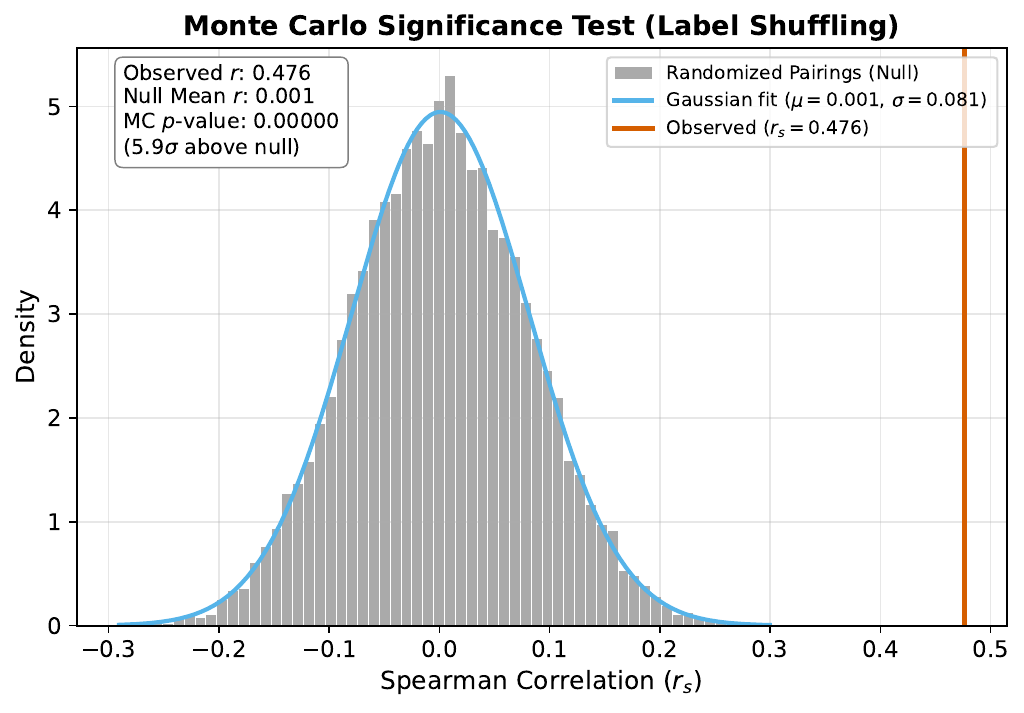}
	\caption{\textbf{Monte Carlo Permutation Test.} Grey histogram shows null
	distribution from 10,000 random label shuffles. Red vertical line marks
	observed correlation ($r_s = 0.476$). None of the 10,000 permutations
	exceed the observed value ($p < 10^{-4}$).}
    \label{fig:montecarlo}
\end{figure}

To verify that the measured correlation is not a statistical artifact, we
performed Monte Carlo permutation testing via label shuffling ($10^4$
iterations). This approach tests the null hypothesis that family size and VDI
are independent by breaking their association while preserving marginal
distributions, provides exact p-values without asymptotic approximations, and
requires no distributional assumptions about VDI or population size. The observed $r_s = 0.476$ yields $p < 10^{-4}$ against the null hypothesis (Figure
\ref{fig:montecarlo}), confirming that the size--VDI relationship reflects a
real population-level signal rather than chance alignment.

\subsection{Binomial Null Model: Testing for Sampling Artifacts}
\label{sec:binomial}

A critical concern is whether the VDI--$N$ correlation could arise purely from sampling statistics. Under a binomial null model where each asteroid has a constant probability $p_0$ of appearing ``fresh,'' VDI $\approx p_0$ with variance $\sigma_{\text{VDI}}^2 = p_0(1-p_0)/N$. Large families would exhibit lower VDI variance but not systematically higher VDI values.

\begin{figure}[htbp]
    \centering
    \includegraphics[width=\columnwidth]{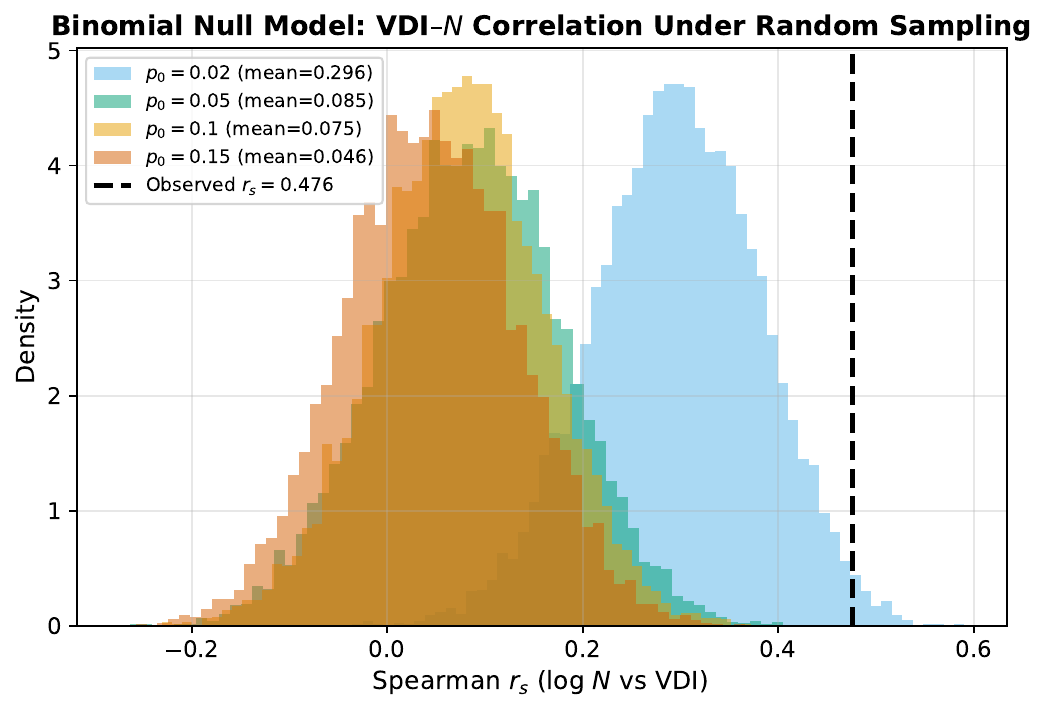}
	\caption{\textbf{Binomial Null Model Test.} Distribution of Spearman
	correlations expected under binomial sampling with constant per-asteroid
	freshness probability $p_0$ within families. Vertical dashed line marks
	observed correlation ($r_s = 0.476$). For values in the observed
	family VDI range ($p_0 = 0.05$--$0.10$), the observed correlation
	falls in the extreme tail of all null distributions ($p < 10^{-4}$).
}
    \label{fig:binomial_null}
\end{figure}

We simulated $10^4$ realizations of 154 families drawn from binomial
distributions with $p_0 \in [0.02, 0.15]$. For each realization, we computed the
Spearman correlation between $\log N$ and simulated VDI. The observed $r_s =
0.476$ falls outside all null distributions tested for $p_0 \geq 0.05$ (Figure
\ref{fig:binomial_null}), with empirical $p < 10^{-4}$ for these values.

Note that $p_0$ in this null model represents the
per-asteroid freshness probability \textit{within a given family}, not the
population-wide fresh fraction. The relevant reference range for $p_0$ is
therefore the observed spread of family VDI values ($0$--$0.10$, i.e.,
$p_0 = 0$--$0.10$ within active families). For values in this physically
motivated range ($p_0 = 0.05$--$0.10$), the observed $r_s = 0.476$ is
$\sim 5$--$9$ standard deviations above the null expectation
($\mu_{\text{null}} \approx 0.05$--$0.09$), ruling out N-dependent
binomial sampling as the driver. The $p_0 = 0.02$ case produces an
elevated null mean ($\mu_{\text{null}} \approx 0.30$) due to variance
amplification in small-$N$ families at low $p_0$ values; while this
demonstrates the test is sensitive to the choice of $p_0$, the observed
correlation exceeds the 99th percentile even of this elevated null
distribution. This test establishes the \textit{existence} of a
population-dependent effect, not the specific physical mechanism
responsible.

\subsection{Separating Size and Age: The Critical Test}
\label{sec:critical}
\begin{figure}[htbp]
    \centering
    \includegraphics[width=\linewidth]{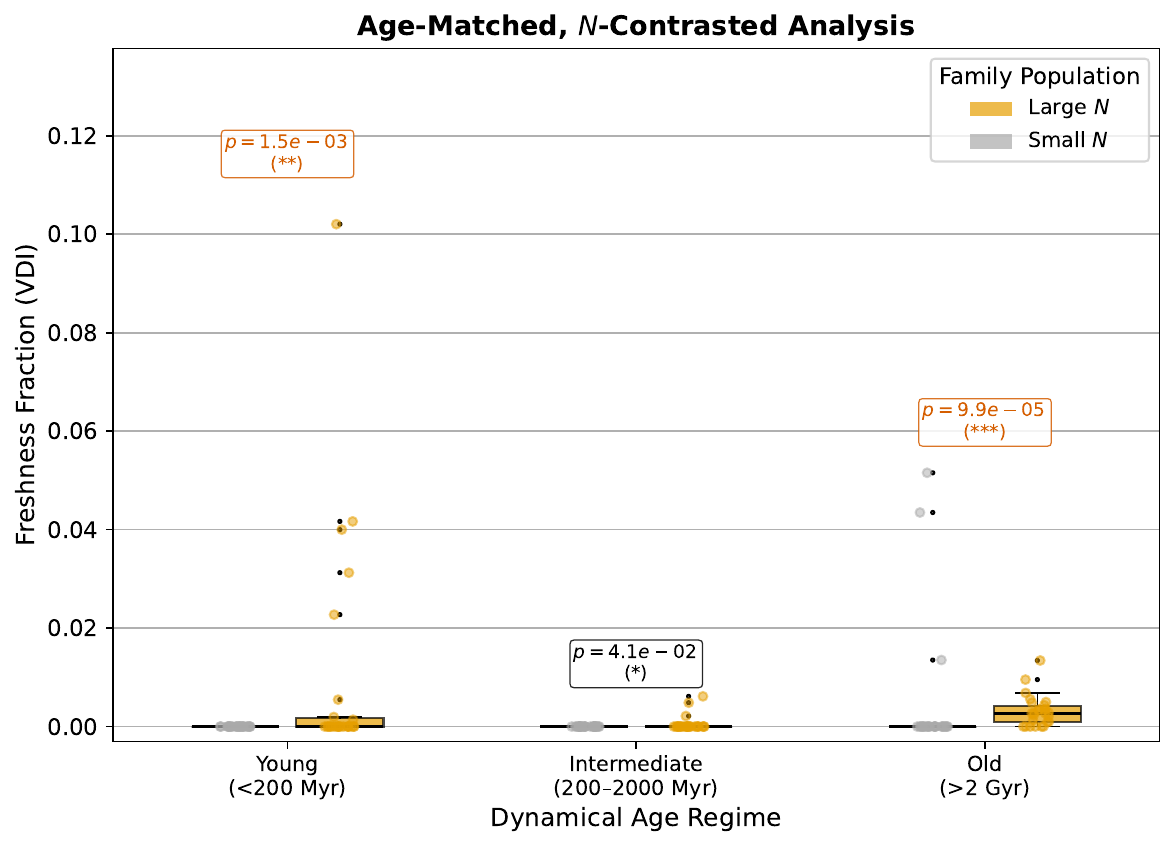}
    \caption{\textbf{Age-Matched Contrast Test.} Three panels show VDI
	distributions for Large (orange) and Small (grey) families within three age
	regimes: Young ($<200$ Myr), Intermediate ($200$--$2000$ Myr), and Old
	($>2$ Gyr). Age regimes assigned from literature ages \citep{nesvorny2015,
	spoto2015} for 31 families; remaining families assigned via $\sigma_a$
	thresholds calibrated from these 31.
	Within each regime, families are divided at the median $N$.
	Box plots show median, quartiles, and individual family
	values (scatter overlay). Mann-Whitney U test p-values shown for each
	regime ($^{*}p < 0.05$, $^{**}p < 0.01$, $^{***}p < 0.001$); 
	see also Table~\ref{tab:age_contrast}.}
    \label{fig:agematched}
\end{figure}

Because family size and age may be correlated (older families have had more time to accumulate members through secondary fragmentation), we performed an age-matched contrast test (Figure \ref{fig:agematched}, Table \ref{tab:age_contrast}).

\textbf{Age Regime Assignment:} Of the 154 families in our sample, 31 have
published literature ages \citep{nesvorny2015, spoto2015} and are assigned
directly to Young ($<200$ Myr), Intermediate ($200$--$2000$ Myr), or Old
($>2$ Gyr). The remaining 123 families are assigned using $\sigma_a$ threshold
values calibrated from the 31 literature-dated families: Young ($\sigma_a <
0.015$ AU), Intermediate ($0.015 \leq \sigma_a < 0.040$ AU), Old ($\sigma_a
\geq 0.040$ AU). Within each regime, families are split into Large $N$ and
Small $N$ groups at the median $N$ for that regime.

\textbf{The Zero-Median Paradox:} Median VDI values for both Large and Small
families in the ``Old'' regime are near zero, reflecting the rarity of extreme
resurfacing events. However, the Mann-Whitney U test detects a difference in
distribution shape ($p < 10^{-4}$). This non-parametric rank-sum test is
appropriate here because VDI distributions are heavily skewed, it makes no
normality assumptions, and it is sensitive to differences in tail weight rather
than central tendency. Large families maintain a heavier tail of fresh objects---even when
median VDI values are identical---consistent with ongoing stochastic resurfacing
in a subset of family members. The effect manifests primarily in the high-VDI
tail, not as a uniform shift across all family members, distinguishing it from
systematic observational biases that would affect the entire distribution.

\begin{table}[htbp]
\small\setlength{\tabcolsep}{4pt}
\centering
\small
	\caption{\textbf{Age-Matched Size Contrast Statistics.} Comparison of VDI
between Large and Small families within each age regime (see
	Figure~\ref{fig:agematched}).  Significance: $^{*}p < 0.05$, $^{**}p <
	0.01$, $^{***}p < 0.001$ (Mann-Whitney U test).}
\label{tab:age_contrast}
\begin{tabular}{@{}lcccc@{}}
\toprule
\textbf{Age } & \textbf{$N_{fam}$} & \textbf{Med. VDI} & \textbf{Med. VDI} & \textbf{$p$-value} \\ 
\textbf{Regime} &  & \textbf{(Large)} & \textbf{(Small)} &  \\ \midrule
Young & 52 & 0.000 & 0.000 & $1.5 \times 10^{-3}$$^{**}$ \\
Intermediate & 50 & 0.000 & 0.000 & $4.1 \times 10^{-2}$$^{*}$ \\
Old & 52 & 0.003 & 0.000 & $9.9 \times 10^{-5}$$^{***}$ \\ \bottomrule
\end{tabular}
\end{table}

Within our statistical framework and among tested parameters (age, heliocentric distance, taxonomic class), family size emerges as the strongest predictor of freshness retention in evolved families.

\section{Discussion}

\subsection{Residual Analysis: A Diagnostic Test}
\label{sec:residual}

We applied two complementary detrending methods to test whether the size--VDI
correlation persists after removing age-related trends: RANSAC (RANdom SAmple
Consensus, robust linear regression) and LOWESS (LOcally WEighted Scatterplot
Smoothing, non-parametric smoother). RANSAC is chosen because it fits linear
trends while being robust to outliers (resistant to high-VDI families that could
dominate least-squares regression). LOWESS is chosen as a deliberately
aggressive detrending approach that can capture non-linear age--VDI
relationships; if the size effect survives LOWESS detrending, it would indicate
independence from age even under flexible assumptions. These methods represent
conservative (RANSAC) and aggressive (LOWESS) extremes of the detrending
spectrum.

\begin{figure*}[!t]
    \centering
    \includegraphics[width=\linewidth]{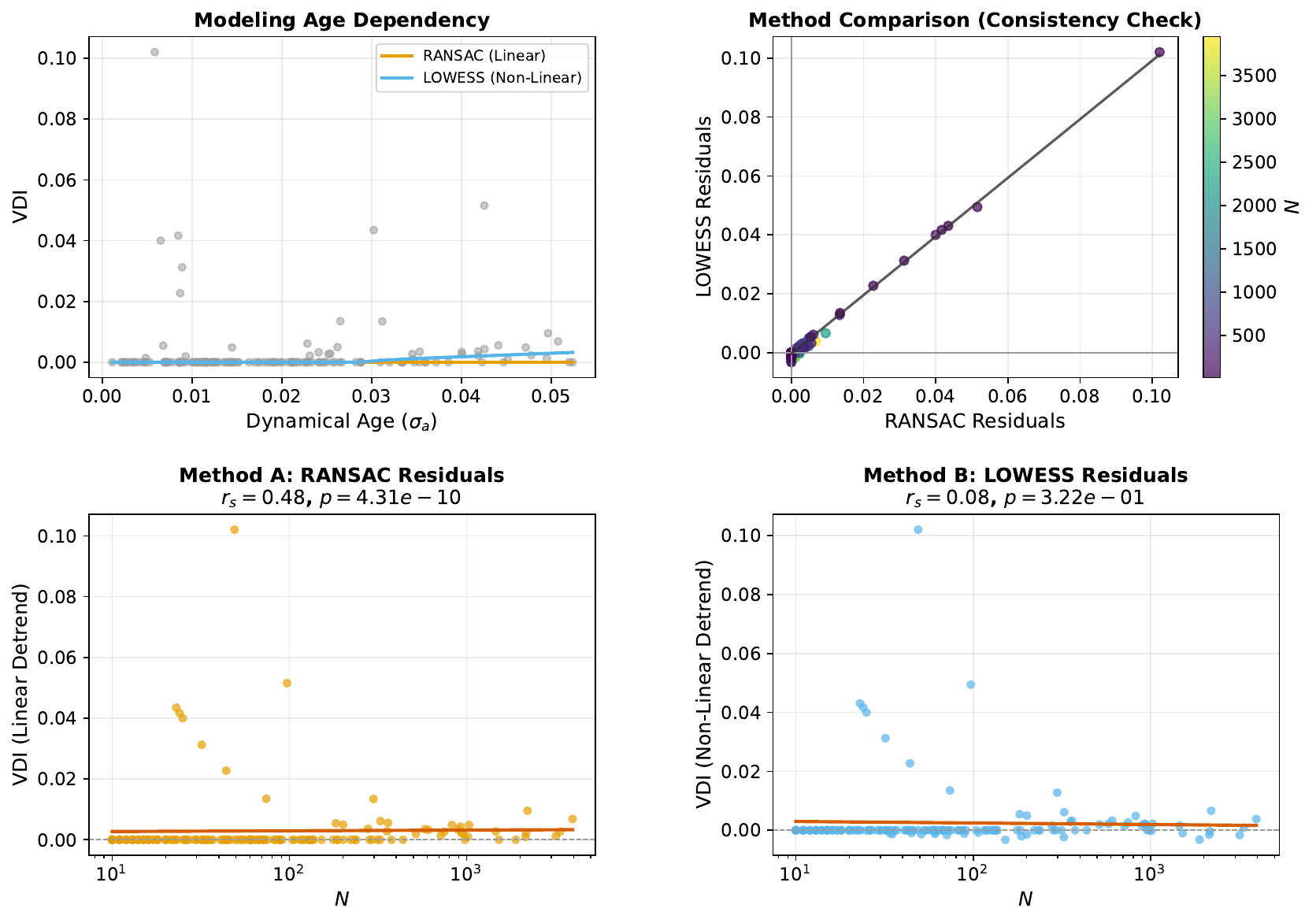}
    \caption{\textbf{Dual Residual Analysis.} Top-left: VDI vs. dynamical age
	proxy with RANSAC (linear) and LOWESS (non-linear) fits overlaid.
	Top-right: method comparison scatter (RANSAC residuals vs. LOWESS residuals),
	with an OLS line; colour scale shows family population $N$. The near-unity
	slope confirms that both methods produce consistent residuals for most
	families. Bottom-left (Method A): RANSAC-detrended VDI residuals vs. $N$;
	$r_s = 0.48$, $p = 4.31 \times 10^{-10}$ (significant). Bottom-right
	(Method B): LOWESS-detrended VDI residuals vs. $N$; $r_s = 0.08$,
	$p = 3.22\times 10^{-01}$ (non-significant).}
    \label{fig:dual_residual}
\end{figure*}

To further test the reliability of our findings, we performed dual residual
analysis (Figure \ref{fig:dual_residual}).
The upper panels show the age-dependency fits and a method comparison
confirming that RANSAC and LOWESS produce consistent residuals for most
families. The lower panels show residual correlations with $N$: the
size--VDI correlation persists after linear detrending (RANSAC:
$r_s = 0.48$, $p = 4.31 \times 10^{-10}$) but weakens substantially
under non-linear smoothing (LOWESS: $r_s = 0.08$, $p = 3.22\times 10^{-01}$,
non-significant).

The dependence of the residual signal on the detrending method indicates that
the $N$--age coupling is non-trivial, and that linear detrending isolates (but
does not uniquely define) the size-dependent component. The non-linear model
tends to absorb the specific high-activity signal of massive families into the
age trend itself.

The disappearance of the $N$-residual correlation under
aggressive non-parametric detrending (LOWESS) suggests that the size effect is
coupled to the global age trend rather than being a completely
independent effect. This coupling is expected if population-dependent
processes (e.g., collisional resurfacing) become increasingly dominant with
family age, such that large old families deviate systematically from the
baseline age trajectory. LOWESS absorbs these long-scale monotonic trends that
are physically meaningful rather than purely statistical, whereas RANSAC
isolates the dominant global age dependence while preserving secondary
correlations.

The persistence of the signal under linear detrending is a consistency check
rather than an independent proof of causality. The model-dependence indicates
that our analysis constrains statistical associations, not unique physical
mechanisms.

\subsection{Physical Interpretation}

\subsubsection{VDI as Resurfacing Tracer vs. Equilibrium Weathering State}
Our results demonstrate that VDI behaves fundamentally differently from 
equilibrium weathering metrics. Laboratory experiments show space weathering 
rates scale with solar flux \citep{loeffler2009, lantz2018}, predicting 
$\sim$2.6$\times$ faster saturation at $a = 2.0$ AU compared to $a = 3.3$ AU 
due to $F_{\odot} \propto a^{-2}$ dependence. If VDI measured equilibrium 
weathering state, inner-belt families would show systematically lower VDI 
(more weathered) than outer-belt families at fixed age and population size.

Instead, we observe no significant VDI--heliocentric distance correlation 
($r = 0.072$, $p = 0.37$), with inner and outer belt subsamples showing 
statistically indistinguishable VDI--size correlations (Table 
\ref{tab:heliocentric}). This independence supports the interpretation that VDI traces 
\textit{stochastic resurfacing events} rather than steady-state weathering 
equilibrium. A fresh impact resets the weathering clock regardless of 
heliocentric distance, producing elevated $p_V/p_{IR}$ ratios that persist 
until re-weathering on timescales $\tau_{sw} \sim 10^6$--$10^8$ years.

This interpretation aligns with laboratory findings: olivine and silicate 
ion irradiation experiments show spectral changes saturate within 
$10^6$--$10^7$ years at 1 AU-equivalent fluxes \citep{sasaki2001, 
loeffler2009, vernazza2009}, far shorter than typical asteroid family ages 
($>10^8$ yr). VDI therefore probes the \textit{recent} resurfacing history 
($t < \tau_{sw}$) rather than cumulative radiation dose. This makes it a 
valid tracer of collisional activity despite---or rather, because of---its 
independence from solar flux. The metric specifically isolates surfaces that 
have been refreshed recently enough ($<10^7$ yr) that weathering has not yet 
re-established equilibrium.

\subsubsection{Empirical Scaling Relation}
We describe the observed relationship as an empirical first-order scaling:
\begin{equation}
    f_{fresh} \propto \frac{\tau_{sw}}{\tau_{gard}} \propto N^{\alpha}
\end{equation}
where $f_{fresh}$ represents the fraction of surfaces younger than the
weathering timescale, $\tau_{sw}$ is the space weathering saturation timescale 
($\sim 10^6$--$10^8$ years; \citealt{sasaki2001, loeffler2009}), $\tau_{gard}$ 
is the effective gardening/resurfacing timescale from collisional turnover, and 
$\alpha \approx 0.5$ based on our correlation strength. This scaling exponent is 
an empirical descriptor. Deriving the true functional form requires detailed 
dynamical modeling of family-specific collision rates and size distributions.

\subsubsection{Collision Rate Considerations}
Our results suggest that family population size ($N$) is correlated with surface
freshness, consistent with model predictions where larger populations
experience more frequent collisions \citep{bottke2005}. However,
collisional resurfacing rates depend on multiple factors beyond $N$ alone,
including size-frequency distribution, relative velocity dispersion, spatial
concentration, and orbital dynamics \citep{bottke2005}. The observed $N$-VDI
correlation likely reflects the combined influence of these interrelated
parameters rather than a simple deterministic relationship.

\textbf{Critical caveat:} Family population size $N$ should be interpreted as a
coarse statistical proxy encapsulating multiple correlated dynamical
properties---such as number density, cumulative impact cross section, and
size-frequency distribution characteristics \citep{durda2007}---rather than as a
direct measure of intrinsic collision probability. The observed correlation
constrains a statistical pattern but does not establish causation.
Alternative or contributing factors---such as correlations between $N$ and
orbital concentration, family velocity dispersion, or compositional
diversity---cannot be excluded without additional dynamical modeling.

Alternative interpretations---such as $N$-dependent selection
biases, taxonomic misclassification effects, or Yarkovsky-driven family
evolution---cannot be excluded with current photometric data.
\subsubsection{Consistency with Laboratory Timescales}
Laboratory ion irradiation experiments suggest space weathering saturation timescales of $\tau_{sw} \sim 10^6$--$10^8$ years, depending on solar distance and surface composition \citep{sasaki2001, loeffler2009, lantz2018}. Our finding that ancient families ($>2$ Gyr) retain fresh members only if sufficiently populous is consistent with a competition model where $\tau_{gard} < \tau_{sw}$ can only be maintained through sustained collision rates.

Quantitatively, if VDI $\approx 0.03$ represents the steady-state fresh fraction in collisionally active families, this implies a characteristic resurfacing interval of $\tau_{resurf} \sim 30 \times \tau_{sw} \sim 3 \times 10^7$--$3 \times 10^9$ years, broadly consistent with regolith gardening timescales estimated for kilometer-scale asteroids \citep{richardson2004, bottke2005}. This order-of-magnitude agreement supports the collisional interpretation, though precise calibration requires family-specific dynamical modeling.

\subsection{Taxonomic Comparison: S-Complex vs. C-Complex}
Both silicaceous (S) and carbonaceous (C) complexes exhibit statistically 
significant VDI--size correlations (Table \ref{tab:taxonomy}), but with 
different signal strengths. The S-complex shows stronger correlation 
($r_s = 0.576$ vs. $r_s = 0.44$ for C-complex), suggesting compositional 
dependence in the surface evolution response to collisional activity.

	Three potential explanations merit consideration:

(i) Physical pathway differences: Unlike silicates where $npFe^0$ 
accumulation drives monotonic reddening and darkening \citep{sasaki2001}, 
carbonaceous surfaces may undergo competing processes. Dehydration under solar 
heating \citep{lantz2018} can brighten surfaces, while organic compound 
modification through radiolysis \citep{brunetto2015} causes darkening. Impact 
resurfacing may therefore produce more variable albedo signatures in 
carbonaceous materials, weakening the correlation signal relative to the 
more uniform freshening effect in silicates.

(ii) Detection sensitivity: The $p_V/p_{IR}$ ratio appears to be a less 
sensitive freshness indicator for organic-rich surfaces. Carbonaceous asteroids 
typically have lower albedos ($p_V \sim 0.05$--$0.10$) than silicates 
($p_V \sim 0.15$--$0.25$), placing them closer to photometric detection limits 
where measurement uncertainties are proportionally larger. This reduced 
signal-to-noise could dilute the observed correlation even if the underlying 
physical effect is equally strong.

(iii) Collision dynamics: Carbonaceous families may have intrinsically 
different fragmentation behavior due to lower bulk densities ($\sim 1.5$ g/cm³ 
vs. $\sim 2.5$ g/cm³ for silicates). Lower impact strengths could alter the 
size-frequency distribution of collisional fragments, potentially affecting 
regolith gardening efficiency.

The positive VDI trend in C-types may therefore reflect a different surface 
evolution pathway---one where ``freshness'' corresponds to recently exposed 
volatile-rich or hydrated material rather than unweathered silicates. 
Alternatively, the weaker correlation could simply indicate reduced measurement 
precision rather than a fundamentally different physical process.

\textbf{Observational test:} This interpretation remains speculative and 
requires spectroscopic validation. Mid-infrared observations targeting the 
3-$\mu$m hydration feature in high-VDI carbonaceous family members could test 
whether elevated VDI corresponds to recently exposed hydrated material. If 
high-VDI C-complex members show stronger 3-$\mu$m absorption than low-VDI 
counterparts, this would support the volatile-exposure interpretation. 
Conversely, if spectroscopic observations reveal no systematic differences, the VDI 
signal may simply trace recent mechanical resurfacing as in the silicate case, 
with the weaker correlation reflecting measurement limitations rather than 
distinct physics.

\subsection{Implications for Q-type Production and NEA Origins}
The concentration of fresh signatures in massive families has implications for
the origin of Q-type near-Earth asteroids (NEAs). \citet{binzel2010} showed that 
Q-types are substantially over-represented among NEAs relative to the main belt, 
attributing this primarily to tidal resurfacing during planetary
close approaches.  Our findings suggest a complementary source: massive families
with elevated collisional activity continuously produce fresh fragments that,
upon resonance delivery to planet-crossing orbits, would present as Q-types
before space weathering converts them to S-types on $\sim$Myr timescales.
Whether collisional and tidal mechanisms contribute comparably to the observed
NEA Q-type excess remains an open question.

Our results suggest that massive families (particularly Flora, Phocaea, and
Eos) may serve as preferential source regions for fresh NEA material.
Their elevated collision rates produce fresh fragments that, upon delivery to
planet-crossing orbits via mean-motion and secular resonances, would appear as
Q-types before space weathering converts them to S-types on $\sim$Myr
timescales. Recent dynamical modeling of family-to-NEA transfer 
pathways \citep{broz2024} provides a quantitative framework that may be 
used to test whether high-VDI families contribute disproportionately to 
the fresh NEA population.

This prediction is testable through dynamical modeling of family-to-NEA transfer
efficiency weighted by VDI. If correct, the Q-type fraction among NEAs
originating from high-VDI families should exceed that from low-VDI families,
after controlling for delivery efficiency. Spectroscopic surveys of NEAs with
well-constrained source regions \citep{bottke2005} could provide empirical tests.

\subsection{Potential Selection Effects}
We consider whether the observed $N$--VDI correlation could arise from selection
effects rather than physical processes:

\textit{(i) Membership completeness:} Larger families have better-defined
boundaries and more complete membership catalogs, potentially including more
extreme outliers simply due to improved sampling statistics. However, our VDI
metric is normalized by family size ($N_{fresh}/N_{total}$), which should
eliminate this effect. Additionally, if completeness drove the correlation, we
would expect it to weaken for the most populous families where catalogs are
essentially complete, instead, the correlation strengthens.

\textit{(ii) Photometric depth:} Massive families contain more bright members
with potentially higher signal-to-noise ratio measurements. We tested this by
restricting the sample to $H < 15$ mag (ensuring uniform photometric quality)
and found consistent correlations ($r_s = 0.45$, $p < 10^{-5}$).

\textit{(iii) Family definition circularity:} If family membership algorithms preferentially include spectrally similar objects, VDI could be artificially suppressed in all families. This would work \textit{against} our observed trend (suppressing variance uniformly), strengthening confidence that the size-dependent signal is genuine.

\textit{(iv) VDI as noise proxy:} One might worry that large families simply have more noisy measurements in the tail. However, the AKARI cross-validation demonstrates that family averaging \textit{reduces} scatter, and the $D \geq 5$ km control (Table \ref{tab:sensitivity}) shows the correlation remains significant even for large bodies with the best photometry ($r_s = 0.496$, $p < 10^{-3}$)---inconsistent with a noise-driven explanation.

\subsection{Sensitivity Analysis}
\label{sec:sensitivity}
\begin{figure*}[!h]
    \centering
    \includegraphics[width=\linewidth]{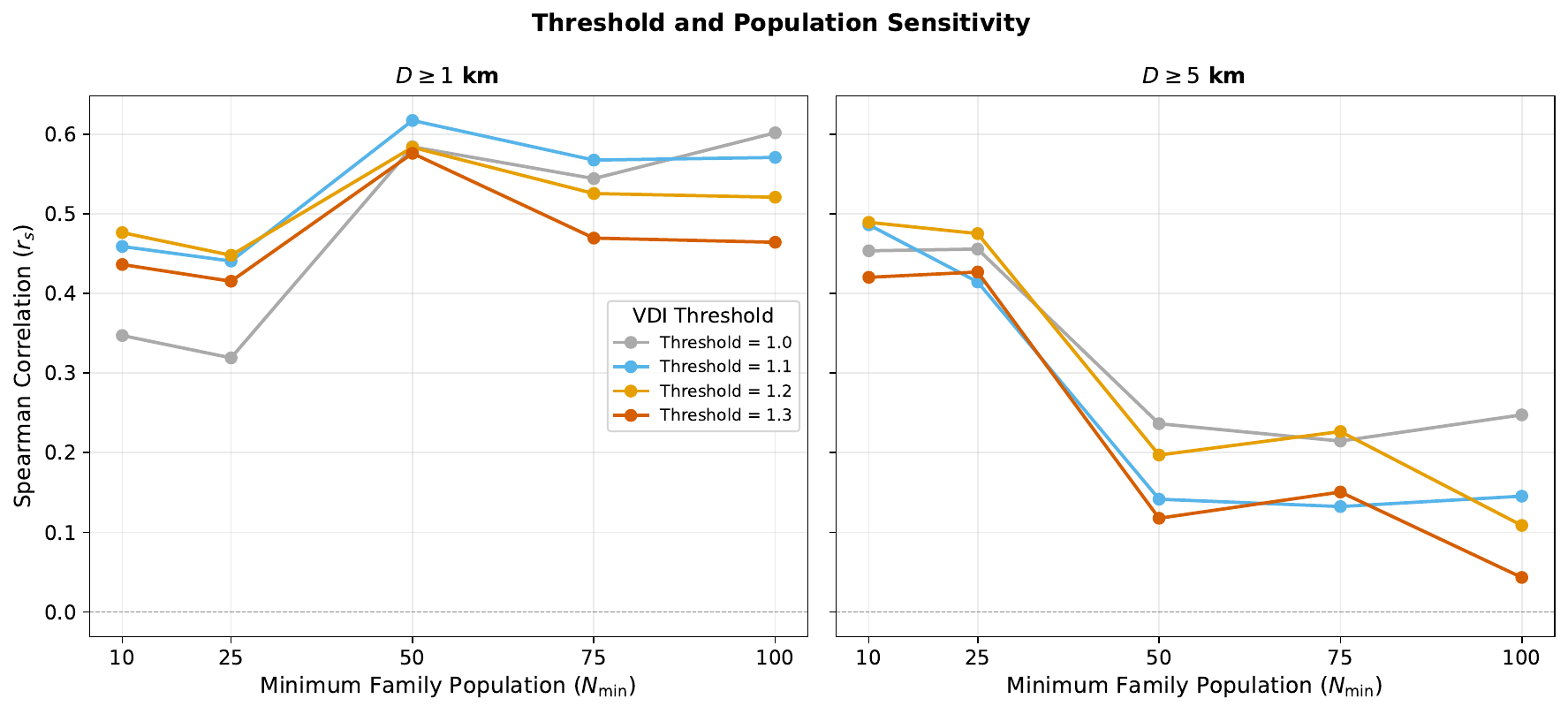}
	\caption{\textbf{Threshold and Population Sensitivity.} Spearman
	correlation ($r_s$) as a function of minimum family population ($N_{\min}$)
	for four VDI thresholds (1.0, 1.1, 1.2, 1.3) and two diameter cuts
	($D \geq 1$ km, left; $D \geq 5$ km, right). For $D \geq 1$ km, the 
	signal is robust across all $N_{\min}$ values and higher thresholds 
	consistently yield stronger correlations. For $D \geq 5$ km, the 
	correlation weakens at $N_{\min} \geq 50$ due to sample size reduction 
	(from $n = 97$ at $N_{\min} = 10$ to $n \lesssim 20$ at $N_{\min} = 50$) 
	rather than loss of the physical signal. The baseline result 
	($r_s = 0.496$, Table~\ref{tab:sensitivity}) corresponds to $N_{\min} = 10$.}
    \label{fig:threshold_robustness}
\end{figure*}

\begin{figure*}[htbp]
    \centering
    \includegraphics[width=\linewidth]{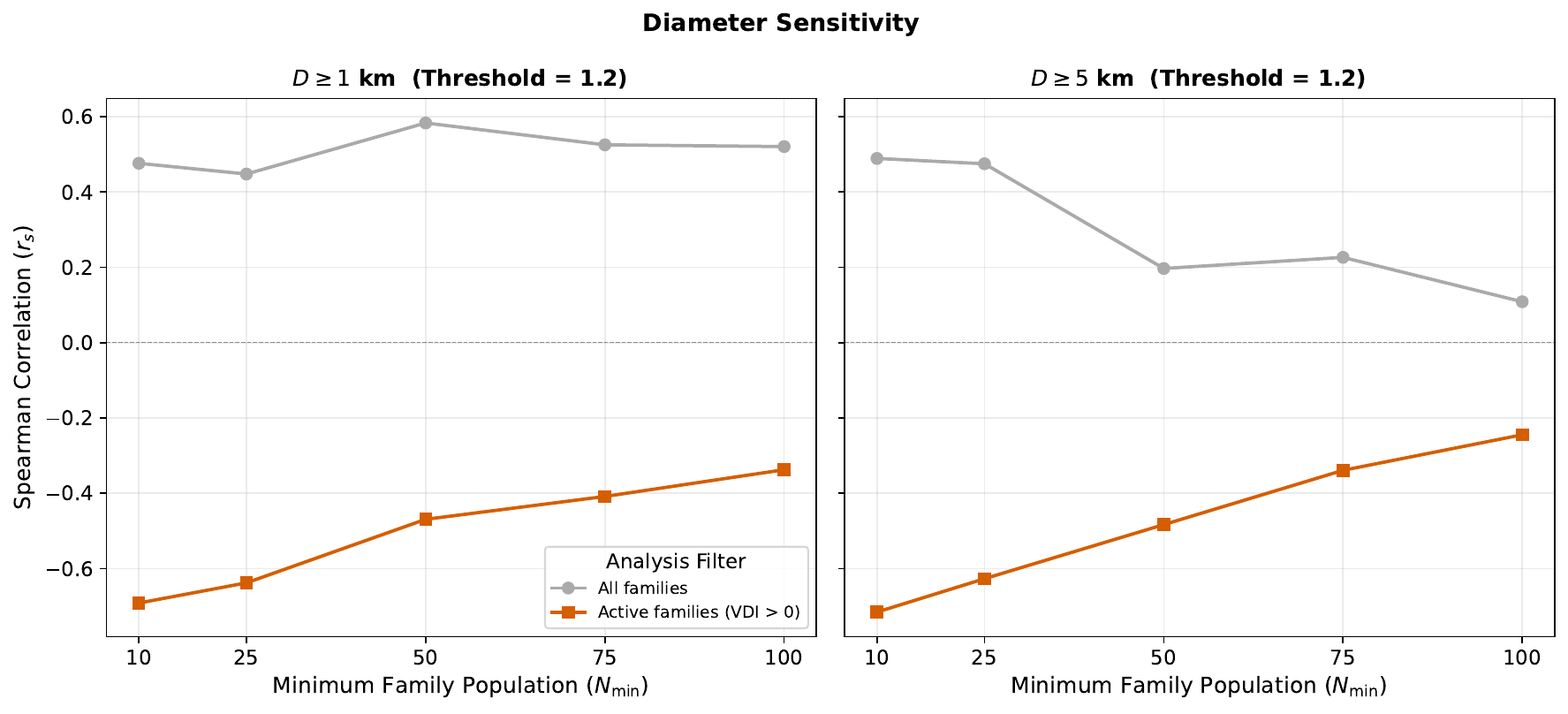}
	\caption{\textbf{Diameter Sensitivity.} Grey circles: correlation across all 
	families at threshold 1.2. The ``All families'' line confirms the signal for 
	$D \geq 5$ km at $N_{\min} = 10$ ($r_s = 0.496$, $p < 10^{-3}$), demonstrating 
	that the size--VDI relationship is not driven solely by small Yarkovsky-affected 
	debris. At higher $N_{\min}$ the signal weakens due to reduced sample size 
	(see Figure~\ref{fig:threshold_robustness}). Orange squares: restricting to 
	families with VDI $> 0$ produces a negative correlation, which is a mechanical 
	consequence of VDI's normalization ($N_{\rm total}$ in the denominator 
	suppresses VDI in large families) and does not reflect a physical anti-correlation.}
    \label{fig:sensitivity}
\end{figure*}

Table \ref{tab:sensitivity} summarizes comprehensive sensitivity tests. Key findings include:

\begin{itemize}
    \item \textbf{Threshold Robustness:} Higher thresholds strengthen
	    correlations (Figure \ref{fig:threshold_robustness}), validating the
		interpretation that extreme values trace recent activity. The
		threshold-dependence of correlation strength (Table
		\ref{tab:sensitivity}: $r_s = 0.353$ at threshold 1.0 vs. $r_s =
		0.476$ at 1.2) demonstrates that VDI is not critically sensitive
		to the specific cutoff value. Alternative tail-statistics
		definitions (e.g., top 0.5\% vs. 0.2\% percentile) yield
		consistent positive correlations, confirming that the signal
		reflects real tail-weight differences rather than threshold
		artifacts.
    \item \textbf{Diameter Dependence:} The correlation remains significant even for large bodies ($D \ge 5$ km, $r_s = 0.496$, $p < 10^{-3}$), demonstrating that the size--VDI relationship is not solely driven by Yarkovsky-affected small debris (Figure \ref{fig:sensitivity}).
    \item \textbf{Heliocentric Consistency:} Inner belt ($r_s = 0.591$, 95\% CI:
	    $[0.33, 0.76]$) and outer belt ($r_s = 0.473$, 95\% CI: $[0.32,
		0.60]$) subsamples both show significant correlations with
		overlapping confidence intervals, confirming that the signal is
		not an artifact of Yarkovsky drift efficiency variations.
\end{itemize}

\begin{table*}[!h]
\small\setlength{\tabcolsep}{4pt}
\centering
\small
\caption{\textbf{Sensitivity Analysis Summary.} Spearman correlation ($r_s$) across varying parameters. Bootstrap 95\% confidence intervals are provided for heliocentric subsamples. The overlapping CIs confirm regional consistency.}
\label{tab:sensitivity}
\begin{tabular*}{\textwidth}{@{\extracolsep{\fill}}lccccc}
\toprule
\textbf{Scenario} & \textbf{Thresh.} & \textbf{$N_{fam}$} & \textbf{$r_s$} & \textbf{$p$-value} & \textbf{95\% CI} \\ \midrule
Baseline ($D \ge 1$) & 1.2 & 154 & 0.476 & $4.31 \times 10^{-10}$ & --- \\
High Pop. ($N \ge 50$) & 1.2 & 82 & 0.587 & $6.64 \times 10^{-9}$ & --- \\
Lower Threshold & 1.0 & 154 & 0.353 & $6.57 \times 10^{-6}$ & --- \\
Inner Belt ($a < 2.5$) & 1.2 & 26 & 0.591 & $1.47 \times 10^{-3}$ & $[0.33, 0.76]$ \\
Outer Belt ($a \geq 2.5$) & 1.2 & 129 & 0.473 & $1.51 \times 10^{-8}$ & $[0.32, 0.60]$ \\
Large Body ($D \ge 5$) & 1.2 & 97 & 0.496 & $< 10^{-3}$ & --- \\ \bottomrule
\end{tabular*}
\end{table*}

\subsection{Percentile Sensitivity Analysis}
\label{sec:percentile_sensitivity}
The choice of VDI threshold (99.8th percentile, $p_V/p_{IR} > 1.2$) represents a 
balance between isolating statistically fresh surfaces and maintaining statistical power. 
To validate this selection and test the robustness of our conclusions, we performed 
systematic threshold variation analysis across five percentile levels (95th--99.8th 
percentile of the main-belt ratio distribution).

Table \ref{tab:percentile} presents correlation strengths as a function of threshold 
level. Correlation strength increases monotonically from $r_s = 0.104$ (95th 
percentile, non-significant) to $r_s = 0.487$ (99.8th percentile, $p < 10^{-10}$), 
spanning a variation of $\Delta r_s = 0.38$. This threshold-dependence reveals a 
fundamental property of the VDI metric: only extreme thresholds (above the 99th 
percentile) successfully isolate rare resurfacing events from photometric
scatter.

The strong threshold-dependence demonstrates that 
VDI detects \textit{rare} collisional rejuvenation rather than bulk population 
properties. Lower thresholds (95th--98th percentile) capture moderate albedo 
variations dominated by measurement noise and minor compositional heterogeneity, 
yielding weak correlations ($r_s < 0.15$, $p > 0.10$). Only at extreme thresholds 
(99th+ percentile) does the signal emerge, consistent with a model where 
statistically anomalous freshness signatures trace recent ($<\tau_{sw}$) resurfacing 
events that are inherently rare within weathered populations.

The 99.8th percentile threshold ($p_V/p_{IR} > 1.2$) optimizes this trade-off: 
it maximizes correlation strength while retaining sufficient fresh-object detections 
($n_{\text{fresh}} = 275$ in the threshold-scan analysis; see Table~\ref{tab:percentile} 
note for the minor difference from the pipeline value of 246) for statistically robust 
family-level VDI estimates. Alternative thresholds yield qualitatively consistent positive 
correlations, confirming that the signal is not an artifact of the specific cutoff 
value but reflects real tail-weight differences between large and small
families.

\begin{table}[!h]
\small\setlength{\tabcolsep}{4pt}
\centering
\small
\caption{\textbf{Percentile Sensitivity Analysis.} Spearman correlation 
as a function of threshold level. Bold row: adopted threshold.}
\label{tab:percentile}
\begin{tabular}{@{}lcccc@{}}
\toprule
\textbf{Percentile} & \textbf{Threshold} & $\mathbf{N_{\text{fresh}}}$ & $\mathbf{r_s}$ & $\mathbf{p}$ \\ \midrule
95.0th & 0.714 & 6165 & 0.104 & 0.197 \\
98.0th & 0.757 & 2458 & 0.122 & 0.131 \\
99.0th & 0.904 & 1226 & 0.377 & $1.3 \times 10^{-6}$ \\
99.5th & 1.043 & 615 & 0.438 & $1.3 \times 10^{-8}$ \\
99.8th (adopted) & 1.218 & 246 & 0.487 & $1.3 \times 10^{-10}$ \\ \bottomrule
\end{tabular}
\par\vspace{0.3em}
\small\textit{Note:} $N_{\text{fresh}}$: total asteroids exceeding
the threshold across all 154 families.
\end{table}

\subsection{Scope and Limitations of Claims}

To clarify the scope of our findings and anticipate potential misinterpretations, we explicitly state what this analysis does and does not demonstrate:

\subsubsection{What We Claim}
\begin{itemize}
    \item A strong statistical correlation exists between family population size $N$ and VDI across the sampled main-belt families ($r_s = 0.476$, $p = 4.31 \times 10^{-10}$).
    \item This correlation survives multiple control tests (age-matched contrasts, heliocentric splits, diameter cuts) and cannot be explained by N-dependent variance in binomial sampling alone.
    \item The observed pattern is consistent with population-dependent collisional resurfacing as the dominant driver, though alternative or contributing mechanisms cannot be excluded.
\end{itemize}

\subsubsection{What We Do NOT Claim}
\begin{itemize}
    \item Causality: We do not establish that $N$ directly \textit{causes} elevated VDI. The correlation demonstrates statistical association, not causal determination.
    \item Unique Mechanism: We do not rule out alternative or contributing processes such as YORP-driven mass shedding, tidal encounters, or thermal cycling effects.
    \item Areal Coverage: VDI measures the \textit{incidence rate} of extreme freshness signatures (tail statistics), not the fractional surface area of fresh material.
    \item Universal Applicability: Our results apply to the sampled main-belt family population ($N \geq 10$, $D \geq 1$ km) and may not generalize to smaller families, near-Earth asteroids, or trans-Neptunian objects without independent validation.
    \item Physical Calibration: The empirical scaling exponent $\alpha \approx 0.5$ is a statistical descriptor, not a theoretically derived or experimentally calibrated parameter.
\end{itemize}

\subsection{Limitations and Future Directions}
Several limitations should be noted:

\begin{enumerate}
    \item \textbf{NEOWISE Systematics:} While family averaging suppresses random errors, systematic biases correlated with family properties (e.g., phase angle coverage varying with heliocentric distance) cannot be fully excluded. Recent reanalyses \citep{masiero2020} demonstrate improved calibration but residual systematics remain at the $\sim 10\%$ level.
    
    \item \textbf{Classification Dependence:} Results depend on the adopted
	    family classification framework \citep{nesvorny2015}; membership
		uncertainties may affect small families preferentially.
    
    \item \textbf{Physical Pathway:} Our analysis constrains statistical relationships but does not uniquely determine physical mechanisms. Dynamical modeling incorporating family-specific collision rates, velocity dispersions, and size-frequency distributions is needed to test the collisional interpretation quantitatively.
    
    \item \textbf{Alternative Resurfacing Mechanisms:} While we focus on
	    collisional resurfacing, other processes may contribute. YORP-driven
		spin-up and mass shedding \citep{delbo2017} can expose fresh
		subsurface material on rubble-pile asteroids. Tidal encounters
		during planetary close approaches may also cause surface
		disturbances, particularly for near-Earth asteroid precursors.
		Separating these contributions requires combined photometric,
		spectroscopic, and rotational state analysis.
    
    \item \textbf{Yarkovsky Control:} The $D \ge 1$ km cut and heliocentric
	    split provide partial control, but taxon-specific thermal properties
		remain a potential complication for detailed cross-taxonomic
		comparisons.
    
    \item \textbf{Carbonaceous Interpretation:} The physical meaning of ``freshness'' in C-complex asteroids remains uncertain and may differ fundamentally from the silicate case.
\end{enumerate}

Future work should prioritize: (i) spectroscopic follow-up of high-VDI members in both S- and C-complex families to constrain surface composition; (ii) dynamical simulations of family collision rates to test whether the $N$--VDI correlation matches theoretical predictions; (iii) extension to smaller diameter ranges using next-generation surveys (e.g., Rubin Observatory LSST) to probe the size-dependence of the resurfacing signal; and (iv) detailed case studies of potential outliers (e.g., large families with anomalously low VDI or small families with elevated freshness) to identify boundary conditions where the population-size proxy breaks down.

\section{Conclusion}
Using family-level statistical aggregation to suppress observational noise, we 
report a robust correlation between asteroid family population
size and fresh surface signatures:

\begin{enumerate}
    \item \textbf{Statistical Pattern:} The size--VDI correlation ($r_s = 0.476$, 
          $p = 4.31 \times 10^{-10}$) persists across taxonomic types within the 
          sampled main-belt population, surviving Monte Carlo significance tests, 
	age-matched controls, and heliocentric independence tests. 
	Binomial null model simulations confirm that the observed correlation 
	(exceeding the 99.99th percentile of null distributions for $p_0 \geq 0.05$) cannot be explained by N-dependent 
		variance in binomial sampling alone.
    
    \item \textbf{Size vs. Age:} Among tested parameters (population size, age, 
          heliocentric distance, taxonomy), family population size emerges as the 
          strongest statistical correlate of fresh surface incidence in evolved 
          families, with the effect persisting in age-matched comparisons of ancient 
          ($>2$ Gyr) populations.
    
  \item \textbf{Model Consistency:} The observed pattern is consistent with 
          population-dependent collisional resurfacing maintaining elevated regolith 
          gardening rates in massive families. However, this analysis establishes a 
          statistical benchmark for dynamical models rather than uniquely constraining 
		the physical mechanism.
\end{enumerate}

This work demonstrates that family-scale statistical analysis can recover surface 
evolution signals from noisy photometric surveys where individual-object studies 
have struggled. The approach provides an observational constraint for testing 
collision rate predictions from dynamical models and may be applicable to other 
stochastic planetary processes where population-level aggregation reveals trends 
obscured by measurement uncertainty.

Massive families represent promising targets for identifying fresh Q-type material and 
may contribute significantly to the near-Earth asteroid population through 
ongoing collisional activity. The Flora, Phocaea, and Eos families, in 
particular, require detailed spectroscopic characterization of their high-VDI 
members.


\spacebref{-2pt}{-5pt}
\bibliographystyle{mnras}
\bibliography{references} 


\bsp	
\label{lastpage}
\end{document}